\documentclass{ws-mpla}
\usepackage[super]{cite}
\usepackage{graphicx}
\usepackage{xspace}
\usepackage{url}
\usepackage{booktabs}
\usepackage{wrapfig,rotating}
\usepackage{sidecap}
\usepackage{amsmath}
\usepackage{amssymb}
\usepackage{mathrsfs}
\usepackage{lineno}
\usepackage{color}

\newcommand{\dzero}     {D\O\xspace}
\newcommand{\wplus}     {$W+$jets\xspace}

\newcommand{\ljets}     {$\ell +$jets\xspace}
\newcommand{\lplustw}     {$\ell +2$\,jets\xspace}
\newcommand{\lplusth}     {$\ell +3$\,jets\xspace}
\newcommand{\lplusfo}     {$\ell +4$\,jets\xspace}
\newcommand{\ttbar}     {$t\bar{t}$\xspace}

\newcommand{\met}       {$\not\!\!E_T$\xspace}
\newcommand{\metMath}       {\not\!\!E_T}
\newcommand{\mm}       {\mathrm}

\newcommand{\mTT}{\ensuremath{m(t\bar{t})}\xspace}

\newcommand{\ptt}{\ensuremath{p_{T}^{\mathrm{top}}}\xspace}

\newcommand{\aetat}{\ensuremath{|y^{\mathrm{top}}|}\xspace}

\newcommand{\pythia}    {\textsc{pythia}\xspace}
\newcommand{\alpgen}    {\textsc{alpgen}\xspace}
\newcommand{\madEvent}     {\textsc{madevent}\xspace}
\newcommand{\mcatnlo}   {\textsc{mc@nlo}\xspace}
\newcommand{\powheg}   {\textsc{powheg}\xspace}

\newcommand\T{\rule{0pt}{2.6ex}}       

\begin{document}


\catchline{}{}{}{}{}

\title{REVIEW OF RECENT MEASUREMENTS IN THE TOP QUARK SECTOR AT THE TEVATRON}

\author{\footnotesize Andreas W.~Jung\footnote{Fermilab.}}

\address{Fermi National Accelerator Laboratory, DAB5 - MS 357, P.O. Box 500,\\
Batavia, IL, 60510,\\
USA\\
ajung@fnal.gov\\
\mbox{FERMILAB-PUB-14-071-PPD}}



\maketitle

\pub{Received (\today)}{Revised (Day Month Year)}

\begin{abstract}
Recent measurements in the top quark sector at the Fermilab Tevatron collider are discussed. Measurements at the Tevatron use up to $9.7$ fb$^{-1}$ of data corresponding to the full data sets recorded by the CDF and \dzero experiments, respectively. This review discusses the most recent measurements of inclusive and differential top quark cross sections in strong and electroweak production of top quarks and related measurements, as well as measurements of angular distributions related to asymmetries in top quark production. Furthermore the current status on the precision measurements of the mass of the top quark is discussed. Where available, combinations of CDF and \dzero results are presented.

\end{abstract}

\ccode{PACS Nos.: {\it 14.65.Ha} {Top Quarks}}

\section{Introduction}
\label{toc:intro}
The top quark, $t$, is the heaviest known elementary particle and was discovered at the Tevatron $p\bar{p}$ collider in 1995 by the CDF and \dzero collaborations \cite{top_disc1,top_disc2} with a mass around $173~\mathrm{GeV}$. Top quark mass measurements together with measurements of the $W$ boson mass and the mass of the recently observed Higgs boson provide a strong self-consistency test of the standard model (SM). The production of top quarks at the Tevatron is dominated by the $q\bar{q}$ annihilation process, contributing 85\% of the total production cross section as opposed to gluon-gluon fusion which contributes only 15\%. In contrast, at the LHC these fractions are approximately reversed. The different initial state allows for complementary and unique measurements at the Tevatron. The top quark has a very short lifetime of $\tau \approx 10^{-25}$ s \cite{width,width2}, which prevents any hadronization process of the top quark, allowing to uniquely observe bare quark properties by measuring properties of the top quarks.\\
Top quarks are produced either as top quark-antiquark pairs (\ttbar) via the strong interaction ($q\bar{q} \rightarrow t\bar{t}$), providing a direct test of quantum chromodynamics (QCD), or as single top quarks via electroweak processes, testing the electroweak theory and providing a direct probe of the CKM matrix element $V_{tb}$. Three different channels contribute to the production of single top quarks: the $t$-channel $(q'g \rightarrow tq\bar{b})$, the $s$-channel $(q\bar{q}' \rightarrow t\bar{b})$, and the associated $tW$-channel $(gb \rightarrow tW)$. The latter has a very small production cross section at the Tevatron, making this channel essentially negligible for most measurements at the Tevatron.\\
The measurements discussed here are the most current ones and use either the dilepton ($\ell \ell$) final state or the lepton+jets (\ljets) final state, where $\ell$ can be an electron or a muon that can also originate from semi-leptonic $\tau$ decays. In the SM branching fraction for top quarks decaying into $Wb$ is almost 100\%. Within the \ljets~final state one of the $W$ bosons (stemming from the decay of the top quarks) decays leptonically, and the other $W$ boson decays hadronically. For the dilepton final state both $W$ bosons decay leptonically.\\ 
This review of the most current results of the Tevatron experiments is loosely related to a talk summarizing measurements at the Tevatron given earlier in 2013 at DESY. It is organized as follows: Section \ref{toc:xsecs} summarizes measurements on inclusive and differential top quark production cross sections and related measurements in the strong and electroweak sector. Section \ref{toc:angular} summarizes measurements of angular distributions related to asymmetries in the production of top quarks and measurements of the correlation of the spin of the top quark. Section \ref{toc:mass} outlines the current status of the precision measurement of the mass of the top quark. Where available, combinations of CDF and \dzero results are discussed as well. A short review like the one presented here can not do justice to the wealth of measurements in the top quark sector at the Tevatron and much more complete information can be found in the publications of each experiment \cite{Tevwebpages,Tevwebpages2}.

\section{Inclusive and Differential Cross Sections}
\label{toc:xsecs}
Measurements of inclusive and differential cross sections deepen our understanding of the theory modeling the production of top quarks. In particular measurements of \ttbar pair production tests perturbative QCD, and provide important information that can improve the simulation of QCD processes. As noted earlier, single top quark production provides tests of the electroweak theory and since new physics can change individual production channels, all production modes of single and pair top quark production are needed to check for possible contributions of new physics. Furthermore, many of what are herein called top quark properties, are in fact differential top quark cross sections of a particular kinematic quantity, for example, the top quark polarization and the forward-backward (charge) asymmetry in \ttbar production at the Tevatron. To identify evidence of new physics in their modeling as provided by perturbative QCD (pQCD) the distributions of these observables needs to be tested as strongly as possible and improved where needed.\\
Theoretical predictions of the \ttbar and single top production processes exist at various orders of perturbation in SM theory. The most recent prediction for \ttbar production is a fully re-summed next-to-next-to-leading log (NNLL) at next-to-next-to-leading order (NNLO) pQCD calculation \cite{nnloTheory}.
\begin{table}[h]
\centering
\tbl{Theoretical predictions for total top quark production cross sections and their uncertainties for the Tevatron and, for comparison, also for the LHC \cite{schannelKido, tchannelKido, kido_stop, twchannelKidoLHC}.}
{\begin{tabular}{lccccc} 
\toprule %
Collider &$\sqrt{s}$ [TeV] & $\sigma_{\mm{t\bar{t}}}$ [pb] & $\sigma_{\mm{s-ch.}}$ [pb] & $\sigma_{\mm{t-ch.}}$ [pb] & $\sigma_{\mm{tW-ch.}}$ [pb] \\ \midrule
Tevatron & 1.96 & $\hphantom{00}7.16$ $^{+0.20}_{-0.23}$ & $1.05 \pm 0.06$ & $\hphantom{0}2.26 \pm 0.12$ & $\hphantom{0}0.30 \pm 0.02$\T \\
LHC & 7 & $172.0\hphantom{0}$ $^{+6.44}_{-7.53}$ & $4.6\hphantom{0} \pm 0.2\hphantom{0}$ & $65.9\hphantom{0} \pm 2.1\hphantom{0}$& $15.7\hphantom{0} \pm 0.6\hphantom{0}$\T \\
\bottomrule
\end{tabular}\label{xsecTheory} }
\end{table}
The total uncertainty from factorization and renormalization scale variation and uncertainties of the parton density distribution function (PDF) is approximately 3.5\% for the Tevatron and approximately 4.3\% for the LHC. Table \ref{xsecTheory} summarizes the predictions for \ttbar and single top quark production at the Tevatron and for comparison also for the LHC center-of-mass energy (using $m_t = 173$ GeV and the MSTW2008NNLO PDF \cite{mstw2008nnlo}).

\subsection{Recent inclusive cross section measurements}
\label{toc:xsec_results}

Measurements of single top quark production in the $s$- and $t$-channel are discussed followed by measurements of inclusive and differential cross section measurements of \ttbar production.

\subsubsection{Measurements of single top quark production cross sections}
\label{toc:xsec_results_stop}
As introduced in Section \ref{toc:intro} there are two dominant channels for producing single top quarks at the Tevatron: the $t$-channel $(q'g \rightarrow tq\bar{b})$ and the $s$-channel $(q\bar{q}' \rightarrow t\bar{b})$. Although the calculated cross sections (see Table \ref{xsecTheory}) are not much smaller compared to the \ttbar cross section, the backgrounds are significantly larger, reducing the signal to background ratio. Simple counting measurements are not possible in this domain and instead more sophisticated algorithms to separate signal and background are applied, for example multivariate analyses techniques \cite{techniques2, techniques3}. The measurements of the single top quark production cross section at the Tevatron rely heavily on the application of those multivariate analyses (MVA).\\

The most recent measurement carried out by \dzero uses 9.7 fb$^{-1}$ of data and selects events in the \ljets channel \cite{d0_stop}. An isolated lepton $\ell$ with a transverse momentum ($p_T$) of at least 20 GeV is required. Furthermore exactly two or exactly three jets are required with $p_T > 20$ GeV and a requirement on the pseudorapidity $\eta$ of the jets of $|\eta| < 2.5$. Further quality cuts to reduce multijet and other background contributions are applied. As mentioned earlier signal events in the $s$- and $t$-channel contain $b$ quarks, allowing further background reduction by imposing requirements on the identification of $b$ quarks in the final state. Jets originating from a $b$ quark are usually identified by means of multivariate discriminants ($b$-tagging) built by the combination of variables describing the properties of secondary vertices and of tracks with large impact parameters relative to the primary vertex. Depending on the channel, either one or two $b$-tagged jets are required.\\

\begin{SCfigure}
\includegraphics[width=0.50\textwidth]{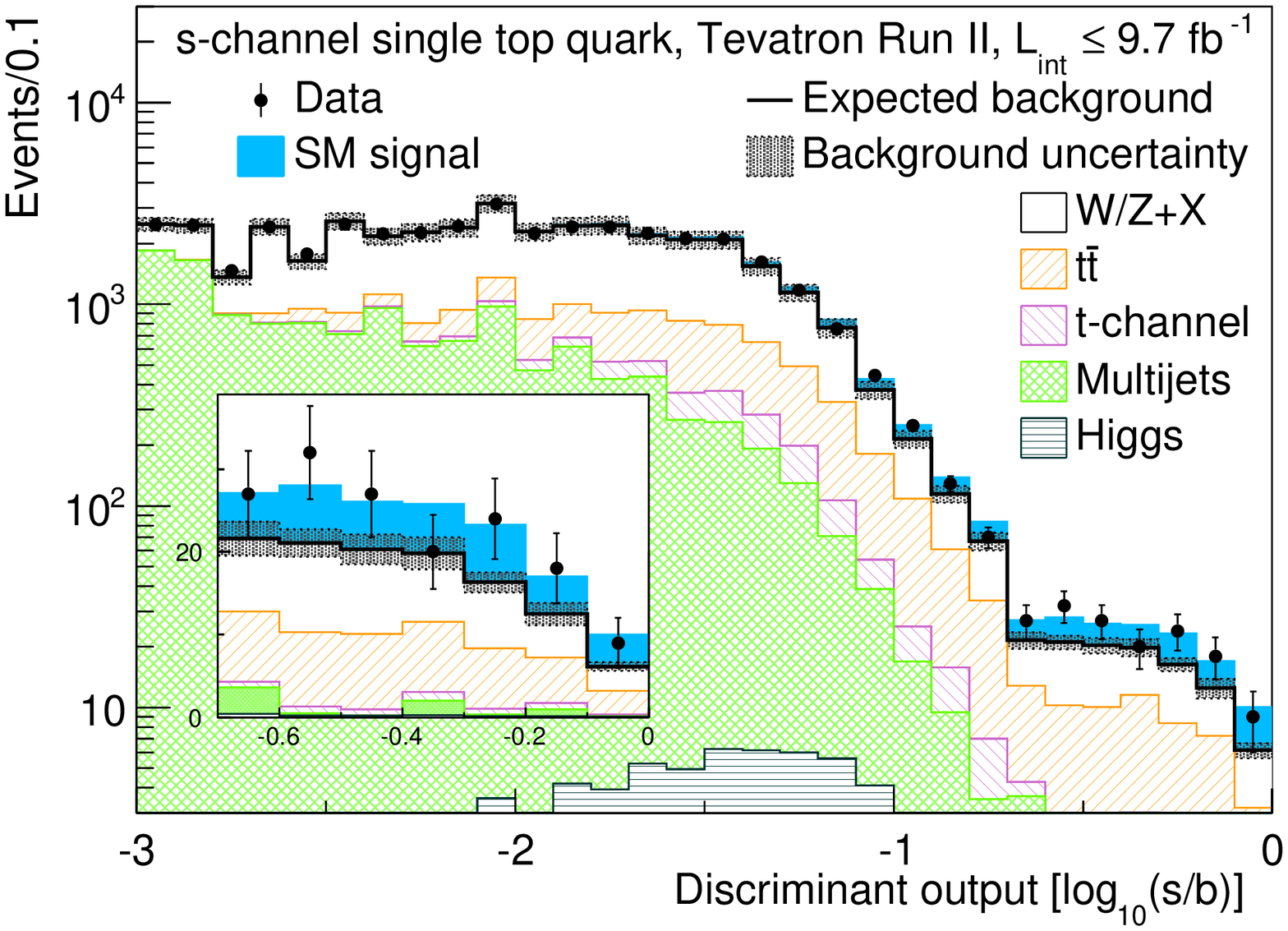}
\protect\caption{\label{fig:tev_stop_obs} Discriminant distribution ranked by the expected signal to background ratio. The $s$-channel single top quark signal and all background contributions are normalized to their expected value. Background uncertainties are indicated by the shaded band on top of the sum of the expected background contributions.}
\end{SCfigure}

The measurement combines three individual MVAs, namely a matrix element analysis \cite{techniques}, a Bayesian neural network \cite {techniques2} and a boosted decision tree \cite{techniques3}, which are combined into a final MVA using a Bayesian neural network.
\begin{figure}[ht]
    \centering
    \includegraphics[width=0.405\columnwidth]{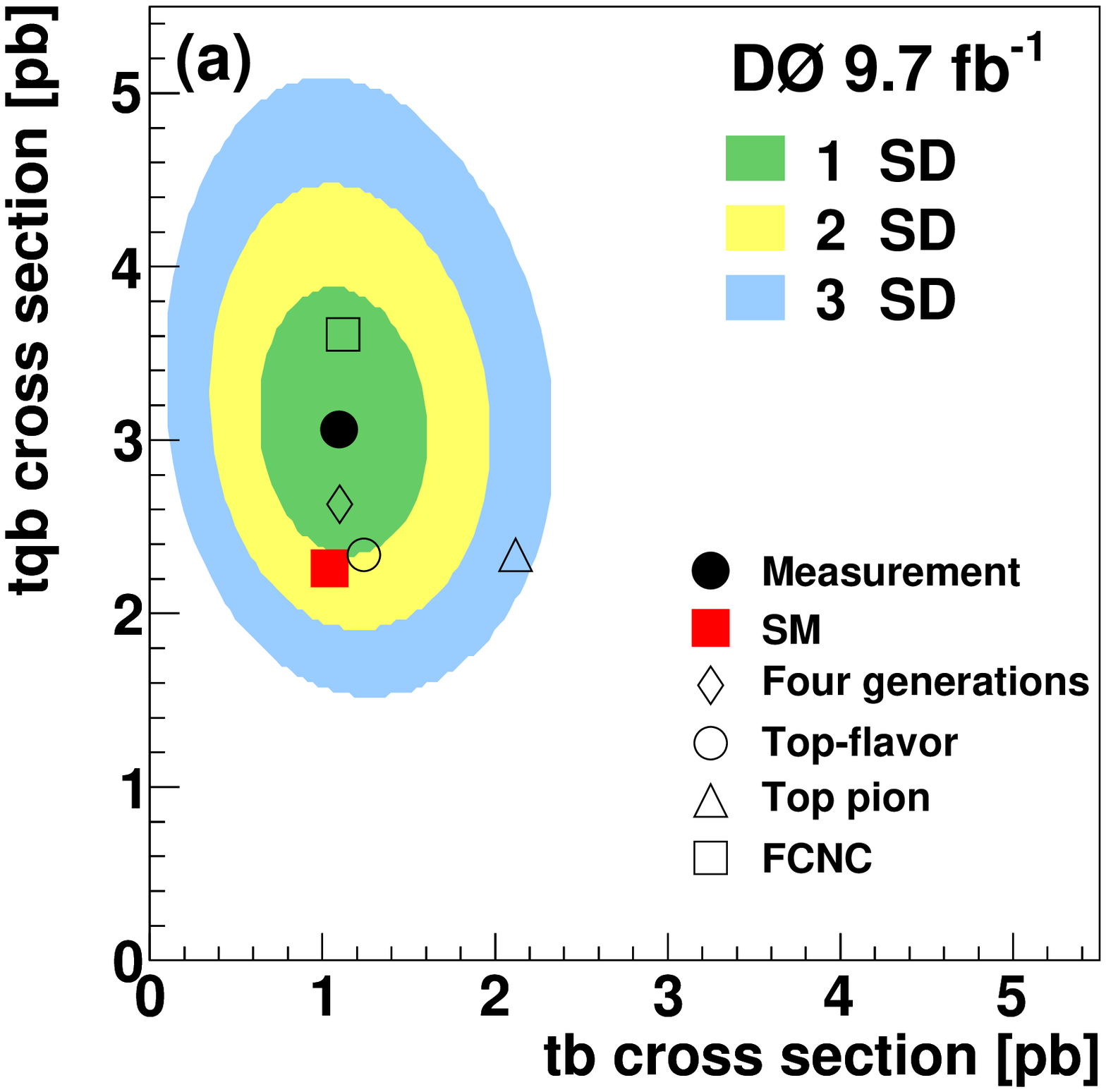}
    \includegraphics[width=0.475\columnwidth]{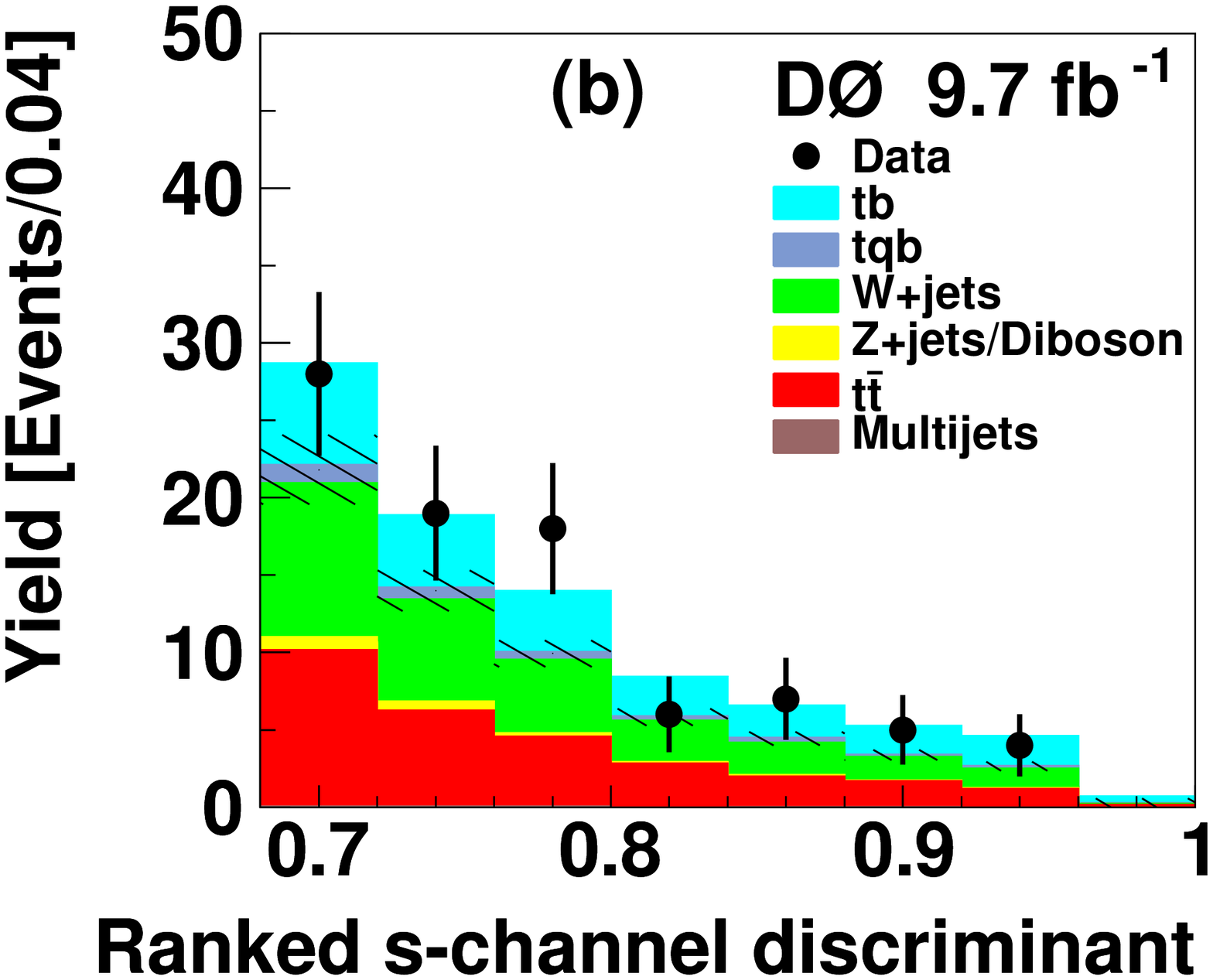}
  \protect\caption{\label{fig:stop_disc} The (a) two-dimensional posterior density with one, two, and three s.d. probability contours compared to various models of new physics altering the $s$- and $t$-channel cross sections, and (b) the signal-dominated region of the ranked $s$-channel discriminant.}
 \end{figure}
Figure \ref{fig:stop_disc}(a) shows the two-dimensional posterior distribution of the simultaneous measurement of the $s$- and $t$-channel cross sections. By employing the two-dimensional discriminant a `multi-purpose' measurement is performed, which allows the measurement of the $s$-, $t$- and $s+t$-channel cross sections in one analysis. Contours are given for one, two and three standard deviations (s.d.) of the measurement uncertainties. The correlation of the three individual MVAs contributing to this measurement is around 75\%.
\begin{figure}[ht]
    \centering
     \includegraphics[width=0.875\columnwidth]{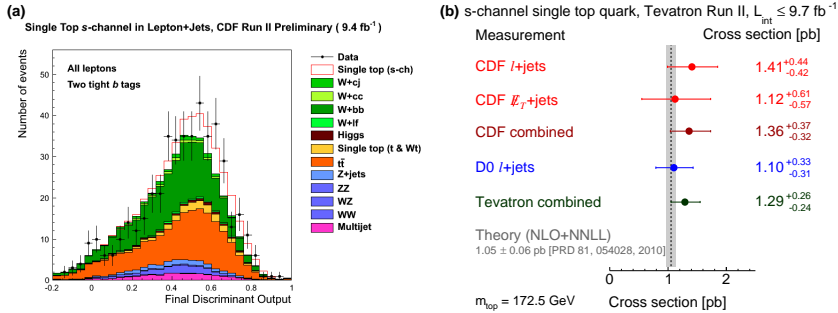}
  \protect\caption{\label{fig:stop_cdf_obs} The (a) final discriminant output distribution of the \ljets channel of CDFs optimized $s$-channel cross section measurement. A (b) summary of all $s$-channel cross section measurements at CDF and \dzero including their combination.}
 \end{figure}
By integrating over the $s$-channel distribution the $t$-channel cross section is measured as $\sigma_{t\mm{-ch.}} = 3.07 +0.53 -0.49$ pb, whereas the $s$-channel measurement yields $\sigma_{s\mm{-ch.}} = 1.10 +0.33 -0.31$ pb after integrating over the $t$-channel distribution. Figure \ref{fig:stop_disc}(b) shows the signal dominated region of the ranked $s$-channel discriminant. The significance of 3.7 s.d. corresponds to the first evidence for $s$-channel production of single top quarks achieved by \dzero. The combined $s+t$-channel cross section is measured to $\sigma_{s+t\mm{-ch.}} = 4.11 +0.59 -0.55$ pb and no assumptions are made on the relative contribution of the $s$- or $t$-channel. A direct limit on the CKM matrix element $V_{tb} > 0.92$ at 95\% confidence level (C.L.) is derived from the combined $s+t$-channel cross section measurement.\\
CDF used 9.4 fb$^{-1}$ for an optimized $s$-channel cross section measurement in the \ljets channel \cite{cdf_stop_ljets}. Events are required to have a high energy isolated lepton candidate with $p_T > 20$ GeV, large $\metMath$ of at least 20 GeV and at least three jets with $|\eta| < 2.0$ and $p_T > 20$ GeV. Figure \ref{fig:stop_cdf_obs}(a) shows the output distribution of the final discriminant for events with two $b$-tags. The $s$-channel contribution is indicated by the open histogram and the measured cross section is $\sigma_{s\mm{-ch.}} = 1.41 +0.44 -0.42$ pb. A measurement in the \met$+$jets channel adds another 10\% of lepton phase space coverage and on its own yields a cross section of $\sigma_{s\mm{-ch.}} = 1.10 +0.65 -0.66$ pb \cite{cdf_stop_met}. In combination with the result from the \ljets channel a cross section of $\sigma_{s\mm{-ch.}} = 1.38 +0.38 -0.37$ pb is measured \cite{cdf_stop_combi}, corresponding to a significance of 4.2 s.d., which confirms the evidence observed by \dzero. Figure \ref{fig:stop_cdf_obs}(b) shows all $s$-channel cross section measurements by CDF and \dzero. All are in good agreement with the latest theoretical calculation of $1.05 \pm 0.06$ pb \cite{kido_stop}. The combination of all existing $s$-channel cross section measurements at the Tevatron yields a cross section of $\sigma_{s\mm{-ch.}} = 1.29 +0.26 -0.24$ pb with a significance of 6.3 s.d. corresponding to the first observation of $s$-channel single top quark production. Figure \ref{fig:tev_stop_obs} shows the discriminant distribution ranked by expected signal to background ratio. The $s$-channel signal and all background contributions and their uncertainties are normalized to the expected value.
The insert shows the signal dominated region, where the data are much better described by the sum of signal and background as opposed to only the background contributions.

\subsubsection{Measurements of \ttbar production cross sections} 
CDF uses all available data corresponding to $8.8~\mathrm{fb^{-1}}$ in the dilepton decay channel to measure the \ttbar production cross section \cite{cdf_dilepton}. The data is selected by requiring exactly two leptons and the accompanying missing transverse energy \met originating from the non-reconstructed neutrinos from the leptonic decays of the two $W$ bosons.
\begin{SCfigure}
\includegraphics[width=0.70\textwidth]{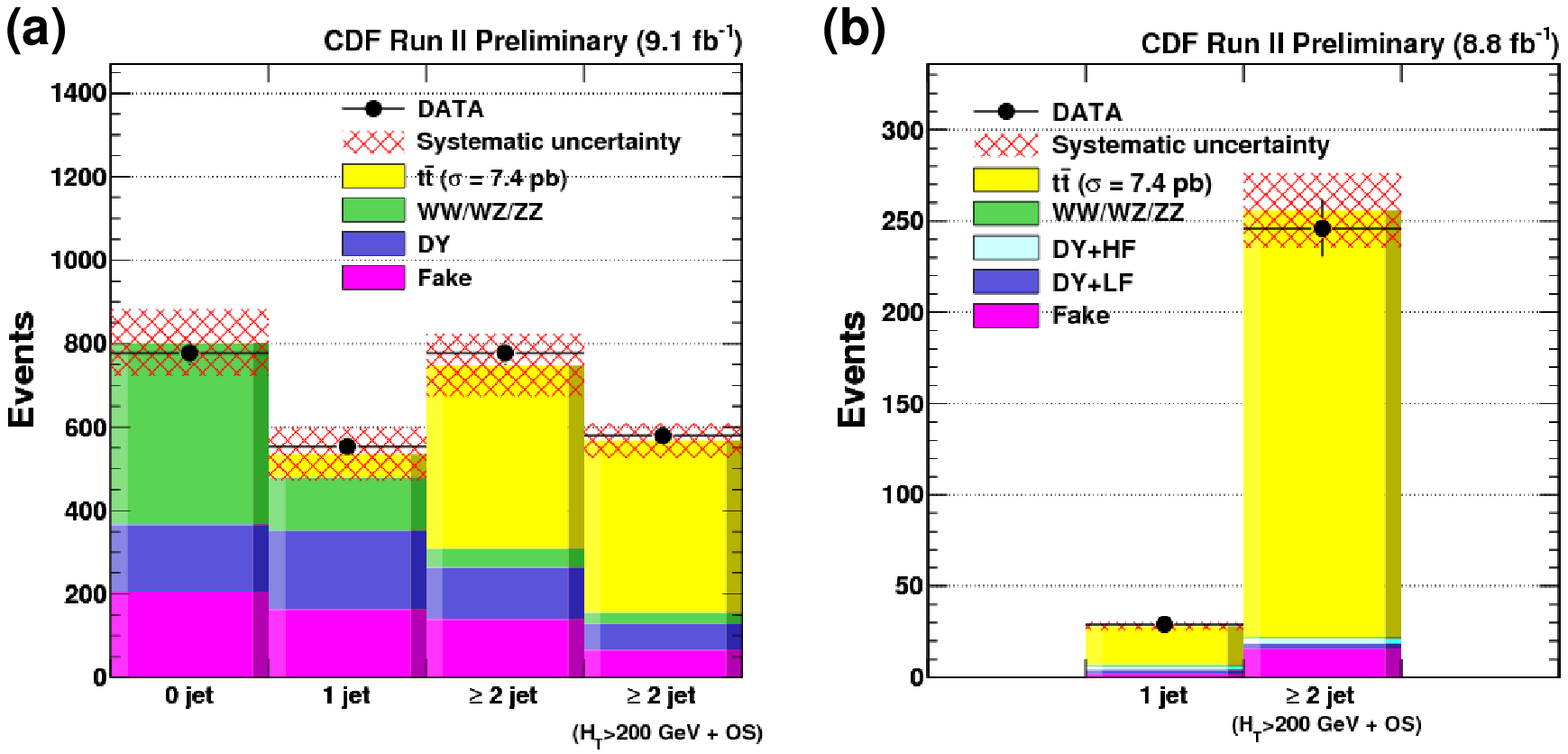}
\protect\caption{\label{fig:cdf_dilepton} Selected data in the dilepton channel for the (a) pre-tag or (b) $b$-tagged case compared to the background contributions. Systematic uncertainties are indicated by the bands on top of the sum of signal and background contributions.}
\end{SCfigure}
Leptonic decays of $\tau$s are included, whereas hadronic ones are not considered here. At least one isolated electron with $E_T > 20$ GeV is required. Muons are required to have at least $p_T > 20$ GeV. Furthermore at least two jets with $p_T > 15$ GeV and pseudorapidity $|\eta| < 2.5$ are required, with at least one $b$-tagged jet. Figure \ref{fig:cdf_dilepton} shows the selected data in the (a) pre-tag or (b) $b$-tagged case. The total cross section, see Eq.~(\ref{eqn:xsecDef}), assuming $m_t = 172.5$ GeV is measured from this $b$-tagged event selection to be $\sigma_{\mm{tot}}(t\bar{t}) = 7.09 \pm 0.49 (\mm{stat}) \pm 0.52 (\mm{sys}) \pm 0.43 (\mm{lumi})$ pb. The systematic uncertainty is dominated by the modeling of the $b$-tagging and the total uncertainty for this measurement is 12\%. If no requirement on $b$-tagging is applied a cross section of $\sigma_{\mm{tot}}(t\bar{t}) = 7.66 \pm 0.44 (\mm{stat}) \pm 0.52 (\mm{sys}) \pm 0.47 (\mm{lumi})$ pb is measured. Table \ref{tab:xsecTevatronSummary} shows a comparison to other CDF measurements in the \ljets channel \cite{cdf_ljetsann,cdf_ljetssvx} and to \dzero measurements described later in this section. The CDF measurements are in good agreement with the most recent pQCD prediction at NNLO, which yields a cross section of $\sigma_{\mm{tot}}(t\bar{t}) = 7.24 ^{+0.23}_{-0.27} (\mathrm{scales} \oplus \mathrm{pdf})$ pb (see Table \label{xsecTheory}).\\

In case of \dzero two recent measurements of the \ttbar cross section are available. The measurement in the dilepton channel corresponds to 5.4 fb$^{-1}$ of integrated luminosity, details of the event selection are given in Ref.\ \cite{d0_dilepton}. The discriminant distribution for identifying jets stemming from $b$ quarks in the four event categories defined by the lepton type and number of jets ($e\mu + 1$ jet, $e\mu + 2$ jet, $ee+2$ jets, and \mbox{$\mu\mu+2$ jets}) is used to maximize a likelihood function for the normalization of the total \ttbar production cross section.
\begin{figure}[h]
\centering
\includegraphics[width=0.975\textwidth]{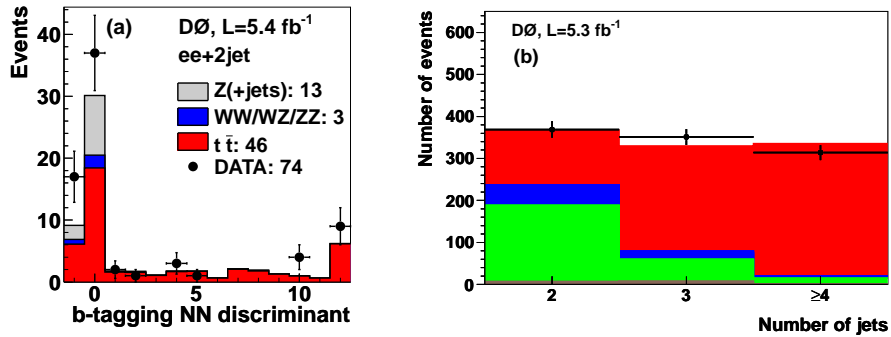}
\protect\caption{\label{fig:d0_dilepton} The $b$-tagging discriminant output is shown in (a) for the $ee + 2$ jets sample, where the expected \ttbar cross section is normalized to $7.45$ pb. The sample composition of the selected \ljets data as a function of the jet multiplicity for (b) more than $1$ $b$-tagged jet with the signal \ttbar contribution in red, \wplus contributions in green, multijet contribution in brown and the sum of all other backgrounds in blue.}
\end{figure}%
An example $b$-tagging discriminant distribution is shown in Fig.~\ref{fig:d0_dilepton}(a), where the expected \ttbar cross section is normalized to $7.45$ pb. The cross section is measured to be $\sigma_{\mm{tot}}(t\bar{t}) = 7.36 ^{+0.90}_{-0.76} (\mm{stat} + \mathrm{sys})$ pb.\\
The measurement in the \ljets channel uses $5.4$ fb$^{-1}$ of data and details on the event selection can be found in the corresponding reference \cite{d0_ljets}. The selected data are used for a combined measurement using $b$-tagging and kinematic information divided into separate channels by number of $b$-tags and jets. Figure \ref{fig:d0_dilepton}(b) shows the sample composition of the selected data as a function of the jet multiplicity requiring more than $1$ $b$-tagged jet. As one expects the background contributions rise towards lower jet multiplicity and the \ttbar contribution rises strongly with number of jets (and also with number of $b$-tags). The cross section is measured to be $\sigma_{\mm{tot}}(t\bar{t}) = 7.78 \pm 0.25(\mm{stat}) \pm ^{0.65}_{0.58}(\mm{sys + lumi})$ pb, which is in good agreement with the theory prediction and other measurements by \dzero or CDF. \dzero is updating the measurements in the \ljets and dilepton channel to the full data sample recorded during Run II.\\
A comparison of the most current measurements at the Tevatron including those discussed above is presented in Table \ref{tab:xsecTevatronSummary}.
\begin{table}[h]
\centering
\tbl{Summary of measurements of the inclusive \ttbar production cross sections and their uncertainties at the Tevatron, including the latest Tevatron combination.}
{\begin{tabular}{@{}lclc@{}}
\toprule %
Measurement & ${\mathscr{L}}$ [fb$^{-1}$] & $\sigma_{\mm{tot}}(t\bar{t})$ [pb] & total rel.\,unc.~[\%] \\ \colrule
CDF ($\ell \ell$, $b$-tag) & 8.8 & $7.09 \pm 0.49\,(\mm{stat}) \pm 0.67\,(\mm{sys})$       & 12.0  \\
CDF (\ljets \cite{cdf_ljetsann}) & 4.6 & $7.82 \pm 0.38\,(\mm{stat}) \pm 0.40\,(\mm{sys})$ & $\hphantom{0}$7.0\\
CDF (\ljets \cite{cdf_ljetssvx}) & 4.3 & $7.32 \pm 0.35\,(\mm{stat}) \pm 0.61\,(\mm{sys})$ & $\hphantom{0}$9.6\\
\dzero ($\ell \ell$) & 5.4 & $7.36\,^{+0.90}_{-0.76}~~(\mm{stat} \oplus \mathrm{sys})$ & 11.0 \\
\dzero (\ljets, $b$-tag) & 5.3 & $7.78 \pm 0.25\,(\mm{stat}) \pm ^{0.65}_{0.58}\,(\mm{sys})$ & $\hphantom{0}$9.1\\ \hline \T
CDF + \dzero (various) & up to 8.8 & $7.60 \pm 0.20\,(\mm{stat}) \pm 0.36\,(\mm{sys})$ & $\hphantom{0}$5.4\\ \\
Theory: & & & \\ \T
NNLO pQCD \cite{nnloTheory} & NA & $7.24 ^{+0.23}_{-0.27}\,(\mathrm{scales} \oplus \mathrm{pdf})$ & $\hphantom{0}$3.5 \\ \botrule
\end {tabular}\label{tab:xsecTevatronSummary} }
\end {table}
The uncertainties of a single measurement at the Tevatron are significantly larger than the uncertainties of the most current pQCD calculation ($\approx 3.5$\%), and only the combination of all available Tevatron cross section measurements \cite{TevatronCombis} yields an uncertainty closer to the theoretical one. The latest Tevatron combination yields $7.60 \pm 0.20\,(\mm{stat}) \pm 0.36\,(\mm{sys})$ with a relative precision of 5.4\%.\\

CDF also measures the cross section for the production of an additional particle in \ttbar production: $p\bar{p} \rightarrow t\bar{t} + \gamma$ \cite{cdf_photon}. The measurement is based on the \ljets decay channel and the selection follows the CDF standard selection for this channel. The additional photon candidate is required to have $E_T^{\gamma} > 10$ GeV, no track with $p_T > 1$ GeV and at most one track with $p_T < 1$ GeV pointing at the electromagnetic calorimeter cluster; and minimal leakage into the hadronic calorimeter. Photons are identified by means of a $\chi^2$ measure and a tight cut on this quantity largely suppresses the background. The most dominant remaining background contribution originates from misidentification of jets as photons. To estimate this contribution the isolation cuts are not applied and $Z \rightarrow ee$ events are used to extrapolate the isolation shape to the applied cuts at higher isolation values. Taking into account this fake photon contribution and all the other background contributions the expected amount of signal candidate events is $26.9 \pm 3.4$ ($e$- and $\mu$-channels combined). The probability of the background alone to mimic the observed signal of 30 events corresponds to three standard deviations (s.d.). The cross section for $p\bar{p} \rightarrow t\bar{t} + \gamma$ is measured to $0.18 \pm 0.07\,(\mm{stat.}) \pm 0.04\,(\mm{syst.}) \pm 0.01\,(\mm{lumi})$, which is a factor of 40 lower than the inclusive \ttbar cross section. The cross section and its ratio to the \ttbar cross section of $0.024 \pm 0.009\,(\mm{stat.}) \pm 0.001\,(\mm{syst.})$ are in good agreement to the SM prediction.

\subsection{Differential cross section measurements in \ttbar production}
\label{toc:xsec_diff}
A differential cross section as a function of the variable $X$ can be calculated using
\begin{equation}
 \label{eqn:xsecDef}
\dfrac{d\sigma_i}{dX} = \dfrac{N^{\mm{obs}}_i - N^{\mm{bg}}_i}{\epsilon \cdot \mathcal{A} \cdot {\mathscr{L}} \cdot {\cal{B}} \cdot \Delta X_i}~.
\end{equation}
The number of observed data events $N^{\mm{obs}}$ is corrected subtracting the number of expected background events $N^{\mm{bg}}$ and then for the detector efficiency $\epsilon$ and acceptance $\mathcal{A}$. The cross section is determined by dividing by the total integrated luminosity $\mathscr{L}$ that corresponds to the selection requirements, for the branching fraction $\cal{B}$ into the decay channel under consideration, and for the bin width of the particular variable $X$. Any measurement of a cross section relies on MC samples to correct the data for the detector efficiency and also in order to extrapolate from the fiducial cross section to the total cross section. For this purpose all cross section measurements use current theory predictions at leading-order or next-to-leading order pQCD. The process of correcting the data for detector effects (see Eq.~(\ref{eqn:xsecDef})) is commonly called unfolding; various approaches to this stage of the analysis exist. They differ in complexity and range from bin-by-bin correction factors to regularized matrix unfolding methods.\\

\begin{figure}[ht]
     \centerline{
       \includegraphics[width=0.375\columnwidth]{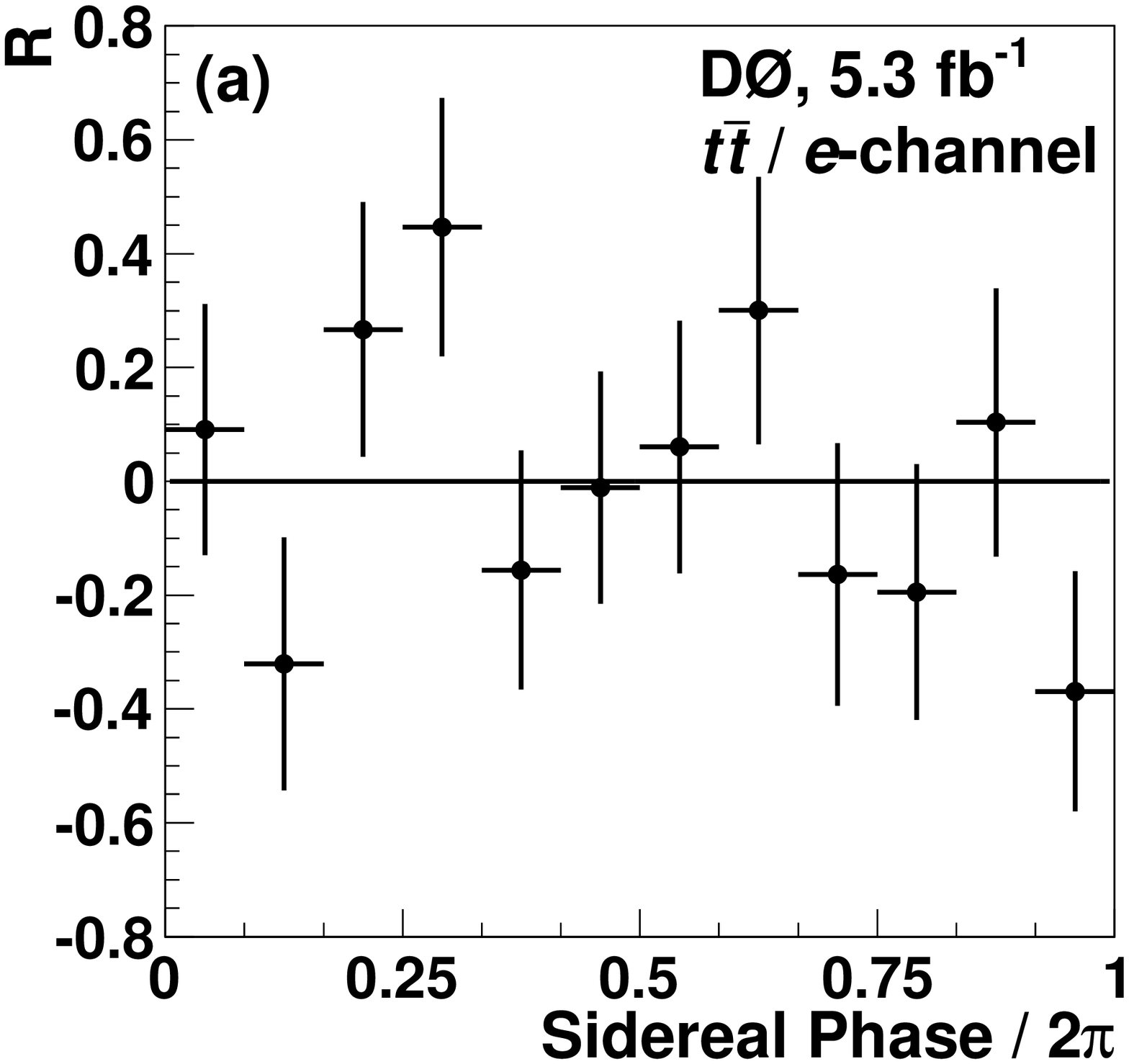}
       \includegraphics[width=0.375\columnwidth]{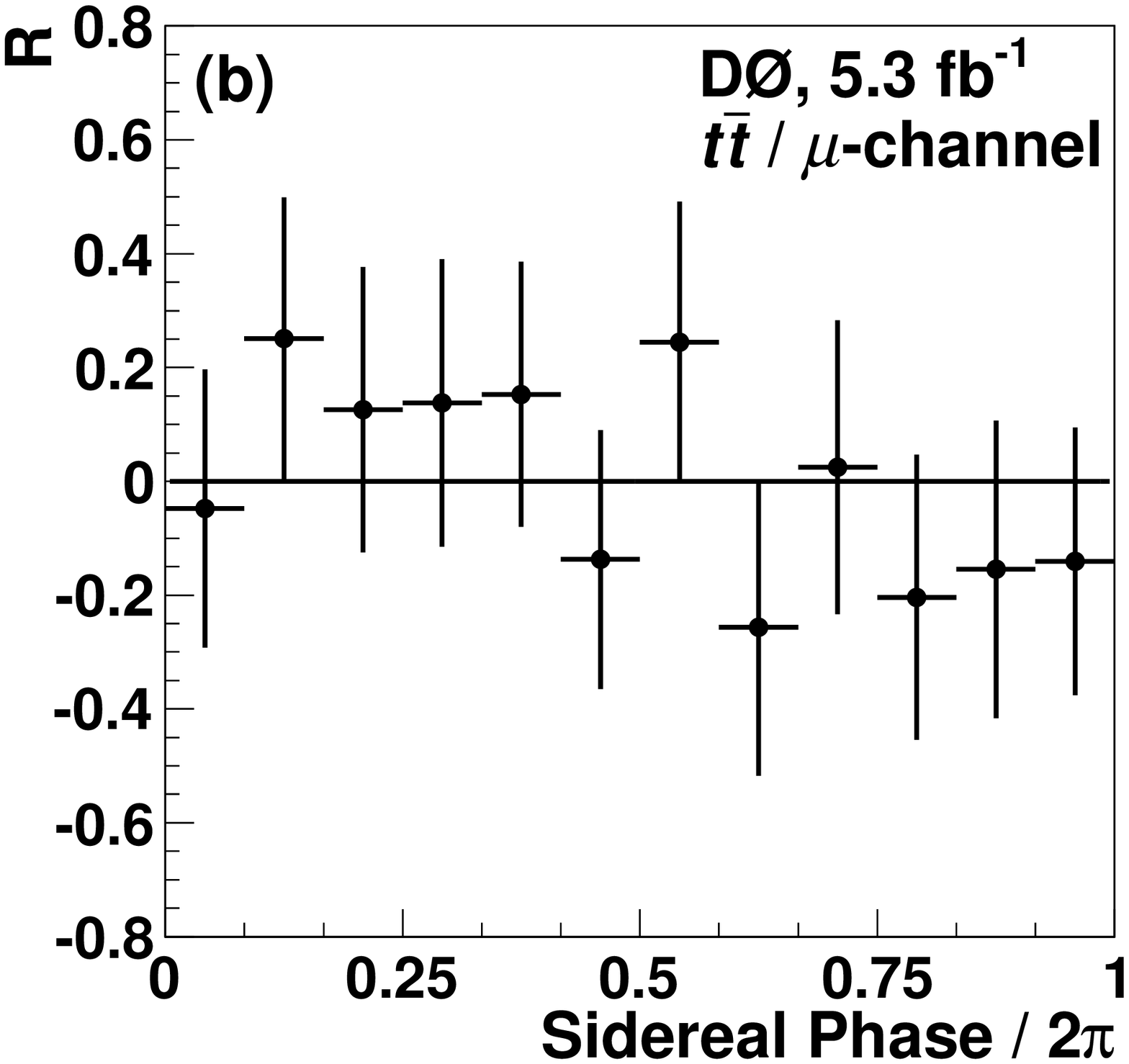}
     }
  \protect\caption{\label{fig:livratio} (a) shows the background and luminosity corrected ratio $R$ as a function of the siderial phase (one siderial day) for \ljets events containing electrons, whereas (b) shows the same ratio $R$ for the muon case.}
 \end{figure}

\dzero searches for a time dependent \ttbar production cross section employing $5.3~\mathrm{fb^{-1}}$ of data \cite{LIVnote}, which is a special type of a differential cross section measurement as a function of the production time. For this analysis \ttbar events in the \ljets~final state are selected with a lepton $(e/\mu)$, at least four jets, exactly one jet identified as a $b$-jet and \met. In addition the analysis relies on the timestamp of the data at production time. The Standard Model Extension (SME) \cite{SMEtheory,SMEtheory2} is an effective field theory and implements terms that violate Lorentz and CPT invariance. The modified SME matrix element adds Lorentz invariance violating terms for the production and decay of \ttbar events to the Standard Model terms. The SME predicts a cross section dependency on siderial time as the orientation of the detector changes with the rotation of the earth relative to the fixed stars. The luminosity-corrected relative \ttbar event rate ($R$) is expected to be flat within the Standard Model, i.e. no time dependency of the \ttbar production cross section. Figure \ref{fig:livratio} shows this ratio as a function of the siderial phase, \mbox{i.e.~1} corresponds to one siderial day.
There is no indication of a time dependent \ttbar production cross section. Instead this measurement sets the first constraints on Lorentz invariance violation in the top quark sector. As the top quark decays before it can hadronize the constraints are also the first ones for a bare quark.\\

The most recent differential measurement uses \ljets events selected using all available data recorded by \dzero experiment \cite{d0note_diff} to study differential top quark cross sections as a function of $p_T$ (\ptt), the absolute value of the rapidity $|y|$ (\aetat), as well as the invariant mass of the \ttbar pair, \mTT. To select \ljets events the following cuts are required: an isolated lepton $(e/\mu)$ with $p_T > 20$ GeV, \met $> 20$ GeV and at least four jets with $p_T > 20$ GeV and $|\eta| < 2.5$. Further cuts are applied to improve agreement of the data with MC and to reject background. To increase the signal purity at least one $b$-tagged jet is required. The sample composition is determined using the discriminant output distribution corresponding to the $b$-tagged jets in the \lplustw, \lplusth and inclusive \lplusfo data.
\begin{SCfigure}
\centering
\includegraphics[width=0.475\textwidth]{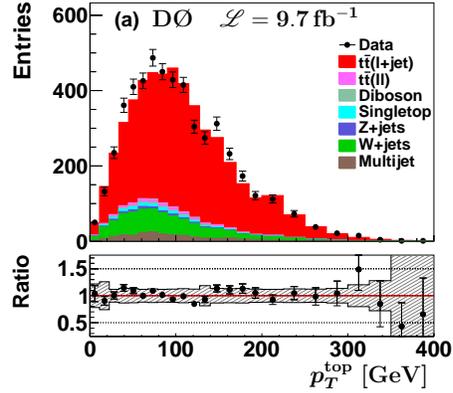}
\protect\caption{\label{fig:ljets97_control} (a) shows the reconstructed \ptt in data after a $\chi^2$ based kinematic event reconstruction compared to the sum of signal and background contributions. The bottom of the panel shows a ratio of the data to the sum of signal and background contributions with the uncertainties indicated by the band. The expected \ttbar cross section is normalized to the measured cross section of 8.27 pb.}
\end{SCfigure}
The established sample composition is employed to measure differential \ttbar cross sections in the \lplusfo bin. To identify the top quarks a kinematic reconstruction is performed, which takes into account experimental resolutions. All possible permutations of objects are considered, while preferentially assigning $b$-tagged jets to $b$-quarks and the chosen solution is the one with the smallest $\chi^2$. As an example the result of the kinematic reconstruction in the case of the \ptt distribution is shown in Fig.~\ref{fig:ljets97_control}(a), where the signal contribution is provided by \mcatnlo using the CTEQ6M PDF \cite{cteq6m}. \\
The \ptt, \aetat and \mTT distributions are background subtracted for all expected contributions and the differential cross sections are determined according to Eq.~(\ref{eqn:xsecDef}), where data is corrected for the detector efficiencies and acceptance as well as the finite detector resolution by means of a regularized matrix unfolding \cite{d0note_diff}.
\begin{figure}[ht]
\centering
    \includegraphics[width=0.825\columnwidth]{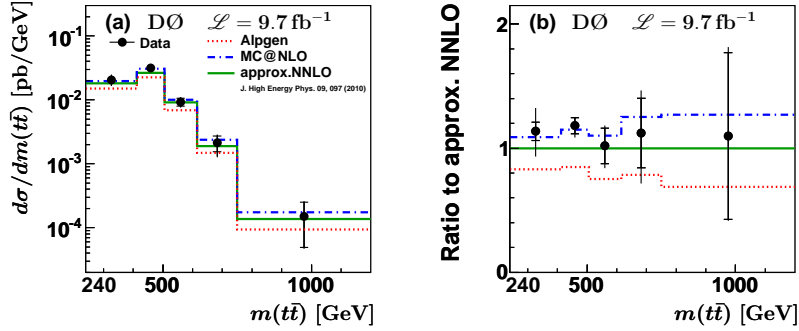}
  \protect\caption{\label{fig:mtt_xsec} Differential cross section as a function of (a) \mTT. The inner error bar corresponds to the statistical uncertainty whereas the outer error bar reflects the total uncertainty. A (b) ratio of the cross section and the predictions to the approximate NNLO \cite{mtt_nnlo} (using MSTW2008nnlo).}
 \end{figure}
Figure \ref{fig:mtt_xsec}(a) shows the differential cross section as a function of the invariant mass \mTT compared to various predictions. By looking at the ratio in Fig.~\ref{fig:mtt_xsec}(b) a reasonable description of the shape by all predictions, with \alpgen being too low in absolute normalization can be seen.

\begin{figure}[ht]
\centering
    \includegraphics[width=0.875\columnwidth]{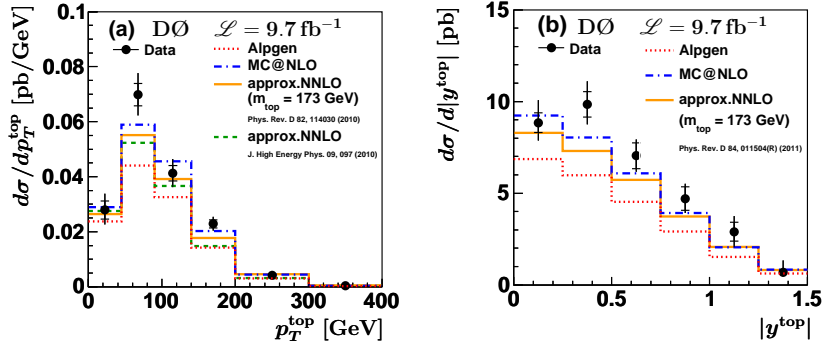}
  \protect\caption{\label{fig:toppt_xsec}Differential cross section data (points) as a function of (a) \ptt and (b) \aetat (two entries per event in each figure) compared with expectations from NLO pQCD (CTEQ6M), from approximate NNLO pQCD calculations (MSTW2008nnlo) \cite{diffKido1,diffKido2}, and from \alpgen (CTEQ6L \cite{cteq6l}). The inner error bar represents the statistical uncertainty, whereas the outer one is the total uncertainty including systematic uncertainties.}
 \end{figure}

Figure \ref{fig:toppt_xsec}(a) shows the differential cross section as a function of \ptt and (b) as a function of \aetat, where the $t$ and $\bar{t}$ contributions are averaged.
Both distributions are described by the various predictions, with some indication of model deficiencies in the low to mid \aetat region where e.g.~the approximate NNLO prediction \cite{diffKido1,diffKido2} is somewhat low compared to data. The measurement in \ptt is in agreement with an earlier measurement by \dzero \cite{d0_toppt} and supersedes it. The total \ttbar cross section is measured to be $\sigma_{\mm{tot}}(t\bar{t}) = 8.27 \pm 0.68\,(\mm{stat}) \pm ^{0.61}_{0.58}\,(\mm{sys}) \pm 0.50\,(\mm{lumi})$ pb, which is somewhat higher than the SM prediction but given the uncertainties still in agreement. A dedicated and more precise inclusive \ttbar cross section measurement in the \ljets channel using the full \dzero Run II data is in progress as well.\\

CDF also measured the differential cross section as a function of \mTT \cite{cdf_mtt} by selecting events with an isolated lepton with a $p_T$ of at least $20~\mm{GeV}$ and a pseudo-rapidity of $|\eta| < 1.1$. A cut on the missing transverse energy of $20~\mm{GeV}$ is applied. Furthermore at least four jets are required with $p_T > 20~\mm{GeV}$ and $|\eta| < 2.0$. Finally at least one jet needs to be identified as a $b$-jet. The hadronic $W$ decay is used to constrain the Jet Energy Scale (JES). \mTT is reconstructed by using the four-vectors of the $b$-tagged jet and the three remaining leading jets in the event, the lepton and the transverse components of the neutrino momentum, given by \met.\\%
\begin{SCfigure}
     \includegraphics[width=0.5\columnwidth]{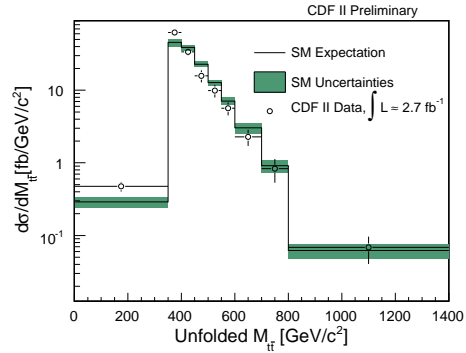}
  \protect\caption{\label{fig:mtt_xsec_cdf} The CDF measurement of the differential \ttbar cross section as a function of \mTT \cite{cdf_mtt} compared to the standard model expectation (CTEQ5L \cite{cteq5l}). The SM uncertainty reflects all systematic uncertainties, except for the luminosity uncertainty in each bin.}
 \end{SCfigure}%
Figure \ref{fig:mtt_xsec_cdf} shows the differential \ttbar cross section as a function of \mTT compared to the standard model expectation using the proton PDF of CTEQ5L with a top quark mass of $175~\mathrm{GeV}$. The SM uncertainty reflects all systematic uncertainties, except for the luminosity uncertainty in each bin. The analysis also measured the inclusive total cross section for \ttbar production to be: $\sigma =  6.9 \pm 1.0\, (\mm{stat.+JES})~\mm{pb}$, which is in good agreement with the latest theoretical predictions \cite{topxsec_theo1,topxsec_theo2,nnloTheory}.

\subsection{Implications}
\label{toc:xsec_implications}
Precise differential cross section distributions provide ample ways of constraining models and sources of new physics. The most direct constraint is set by the \mTT distribution, which is sensitive to the production of resonances decaying into top quarks, like a $Z'$. Another example is the currently puzzling situation at the Tevatron in terms of the enhanced forward-backward asymmetry observed in \ttbar production. The differential top quark cross sections constrain models of new physics, which could explain the effect.\\
CDF employs the tail of \mTT to constrain new physics contributions \cite{cdf_mtt}. The tail is sensitive to broad enhancements as well as to narrow resonances. There is no indication of beyond standard model contributions to the differential cross section. Furthermore the distribution has been used to derive a limit on gravitons which decay to top quarks in the Randall-Sundrum model. The mass of the first resonance is fixed to 600 GeV and gravitons are modeled using \madEvent plus \pythia for modeling the hadronization effects. Values of $\kappa /M_{Pl} > 0.16$ are excluded at the 95\% confidence level.\\

\begin{figure}[htb]
\begin{center}
  \includegraphics[width=0.875\columnwidth,angle=0]{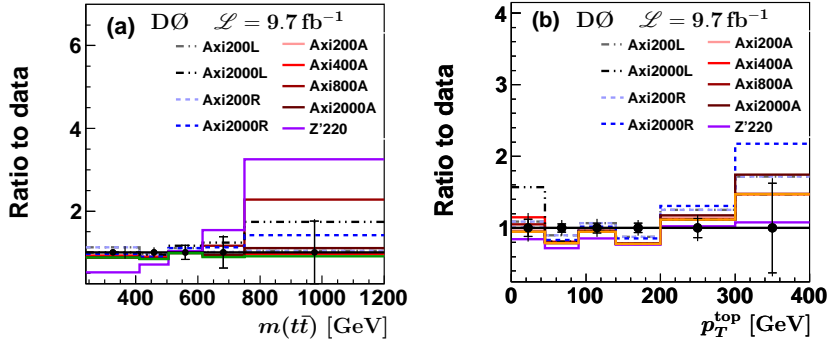}
\end{center}
\protect\caption{Differential cross section distributions as a function of \mTT (top) and \ptt (bottom) compared to various benchmark models of axigluon contributions to the \ttbar production cross section. The QCD prediction at approximate NNLO and the sum with the various axigluon models are compared to the data.}
\label{fig:topxsec_axiModels1}
\end{figure}

As mentioned before an example of the importance of accurate modeling of QCD is given by the deviation observed in the asymmetry measurements in $p\bar{p} \rightarrow t\bar{t}$ production from SM predictions (see Section \ref{toc:angular}). Such a difference could be due to the exchange of a new heavy mediator, e.g., an axigluon \cite{axi1,axi2} that could also enhance the \ttbar cross section. Differential cross sections provide stringent constraints on axigluon models, where instead of the intermediate gluon an axigluon (or a light $Z'$ in alternative models) is produced. These models are constructed to result in the observed asymmetries at the Tevatron (see Section \ref{toc:angular}) and are briefly summarized in Table \ref{tab:chi2IncCorrAxi} in terms of their couplings, masses and corrections to \ttbar cross sections \cite{axigluon}.
\begin {table}[tp]%
\centering %
\tbl{Table of $\chi^2/ndf$ values for data (total uncertainty) versus approximate pQCD at NNLO (scale and PDF uncertainties) and the various axigluon models (scale uncertainty) and one $Z'$ model (scale uncertainty). The nature of the couplings (`$l$' left, `$r$' right and `$a$' axial couplings) is given in the first column, whereas the masses of the new mediators are indicated in the second column in TeV. More details can be found in Ref.\ \cite{axigluon}.}
{\begin {tabular}{lccccc}
\toprule %
               &           & $\sigma_{\mm{tot}}(t\bar{t})$ [pb]  & \mTT $[\chi^2/ndf]$ & \aetat $[\chi^2/ndf]$ & \ptt $[\chi^2/ndf]$ \\ \midrule
Data           &           & $8.27 ^{+0.92}_{-0.91}$ & n.a. & n.a. & n.a.\\
SM      &           & $7.24 ^{+0.23}_{-0.27}$ & 0.98 & 3.71 & 4.05 \\ \\
model   &$m$ [TeV]  & $\Delta \sigma_{\mm{tot}}(t\bar{t})$ [pb] & \mTT $[\chi^2/ndf]$ & \aetat $[\chi^2/ndf]$ & \ptt $[\chi^2/ndf]$ \\ \midrule
$G'(l)$ & $0.2\hphantom{0}$          & $+0.97 \pm 0.06$ & 0.96 & 1.07 & 1.20 \\
$G'(r)$ & $0.2\hphantom{0}$          & $+0.97 \pm 0.06$ & 0.96 & 1.07 & 1.20 \\
$G'(a)$ & $0.2\hphantom{0}$          & $+0.06 \pm 0.04$ & 0.85 & 3.55 & 3.88 \\
$G'(a)$ & $0.4\hphantom{0}$          & $+0.26 \pm 0.04$ & 0.44 & 2.65 & 3.26 \\
$G'(a)$ & $0.8\hphantom{0}$          & $+0.22 \pm 0.04$ & 0.97 & 2.86 & 3.23 \\
$G'(l)$ & $2.0\hphantom{0}$          & $+0.87 \pm 0.15$ & 0.58 & 1.27 & 3.78 \\
$G'(r)$ & $2.0\hphantom{0}$          & $+0.55 \pm 0.06$ & 0.43 & 1.94 & 2.75 \\
$G'(a)$ & $2.0\hphantom{0}$          & $+0.05 \pm 0.06$ & 0.88 & 3.56 & 4.11 \\
$Z'$    & $0.22$                     & $-1.00 \pm 0.06$ & 4.95 & 8.27 & 7.48 \\ \bottomrule
\end {tabular} \label{tab:chi2IncCorrAxi}}
\end {table}
In addition Table \ref{tab:chi2IncCorrAxi} states the $\chi^2/ndf$ values of the measured differential cross sections compared to the various models. The $\chi^2$ takes into account the full covariance matrices of the differential cross section measurement in \ptt, \aetat and \mTT. Models implementing heavy masses are usually in tension with existing data from the Tevatron and the LHC, but it is especially the low mass region where the Tevatron data adds sensitivity. Figure \ref{fig:topxsec_axiModels1}(a) shows a ratio of the various models introduced earlier to the measured differential \ttbar cross section in \mTT. Figure \ref{fig:topxsec_axiModels1}(b) shows the same comparison using the \ptt distribution.
In contrast to the expectation that \mTT is the most sensitive distribution, \ptt (and \aetat) also add sensitivity. In particular the region with high experimental precision (medium values of \ptt) offers sensitivity to these low mass models. The low mass $Z'$ model shows significant tension to the data in all three differential distributions as can be seen from the ratios with the $\chi^2/ndf$ values summarized in Table \ref{tab:chi2IncCorrAxi}.

\section{Measurements of Forward-backward Asymmetries}
\label{toc:angular}
The different initial state makes measurements of angular correlations in $t\bar{t}$ events, like forward-backward asymmetries, at the Tevatron complementary to those at the LHC. Another example not summarized here are spin correlation measurements (more information can be found in Refs.~\cite{Tevwebpages,Tevwebpages2}).\\
Experimentally one needs to distinguish between two approaches to measure these asymmetries: Either the \ttbar pair is fully reconstructed using a kinematic reconstruction \cite{d0_afb,d0_afb_update,cdf_afb} or only a final state particle like the lepton (`lepton based asymmetries') is reconstructed \cite{d0_afb_lep,cdf_afb_lep}. The latter avoids the reconstruction of top-quarks, which is usually more affected by detector resolution and migration effects. In general the size of the \ttbar asymmetries at the Tevatron or at the LHC is related to the relative weight of the quark annihilation channel in the pair production, because events arising from $gg$ initial states are forward-backward symmetric. At the Tevatron the $t$ and $\bar{t}$ rapidity distributions are shifted with respect to each other, which is why the measurement of the forward-backward asymmetry in \ttbar production ($A_{\mbox{{\footnotesize FB}}}$) relies on a measurement of $\Delta y = y_t - y_{\bar{t}}$. The forward-backward asymmetry is defined as
\begin{equation}
A_{\mbox{{\footnotesize FB}}}^{t\bar{t}} = \dfrac{N(\Delta y >0) - N(\Delta y <0)}{N(\Delta y >0) + N(\Delta y <0)}~.
\end{equation}
The charge asymmetry $A_{\mbox{{\footnotesize C}}}$ at the LHC arises from the fact that quarks have on average a larger longitudinal momentum than anti-quarks, which in case of a $pp$ collider leads to a wider rapidity distribution in the case of $t$ production compared to the $\bar{t}$.\\
An additional observable is given by the lepton based asymmetries, which are defined in the following way employing measurements of the charge $q_{\ell}$ and $\eta$ of the leptons:
\begin{equation}
A_{\mbox{{\footnotesize FB}}}^{\ell} = \dfrac{N(q_{\ell} \cdot \eta >0) - N(q_{\ell} \cdot \eta <0)}{N(q_{\ell} \cdot \eta >0) + N(q_{\ell} \cdot \eta <0)}~,
\end{equation}
for the single-lepton asymmetry $A_{\mbox{{\footnotesize FB}}}^{\ell}$ and for the dilepton asymmetry $A^{\ell \ell}$:
\begin{equation}
A^{\ell \ell} = \dfrac{N(\Delta \eta >0) - N(\Delta \eta <0)}{N(\Delta \eta >0) + N(\Delta \eta <0)}~.
\end{equation}
The difference $\Delta \eta$ is given by $\eta_{\ell^+} - \eta_{\ell^-}$ (signs refer to charge of the lepton). The two lepton based asymmetries are correlated, but by combining them a small reduction of uncertainties is observed.

\begin{figure}
  \centering
  \includegraphics[width=0.975\columnwidth,angle=0]{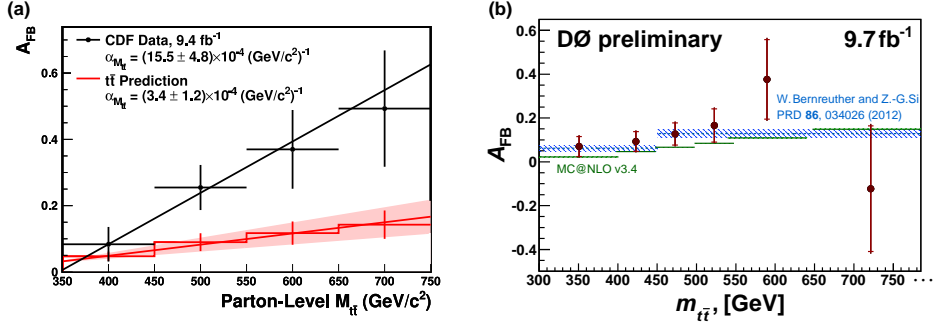}
\protect\caption{\label{fig:cdf_mttafb} The (a) $A_{\mbox{{\footnotesize FB}}}^{t\bar{t}}$ at parton level as a function of \mTT compared to the predicted dependency by \powheg as measured by CDF. A (b) similar measurement by \dzero (see text for more details).}
\end{figure}

\begin{SCfigure}
    \centering
    \includegraphics[width=0.4\columnwidth]{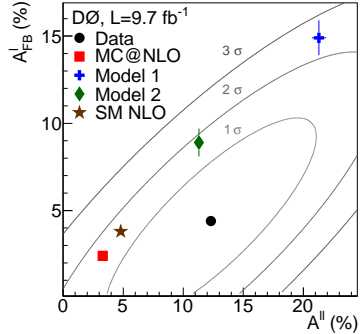}
  \protect\caption{\label{fig:al_all_d0} Extrapolated $A_{\mbox{{\footnotesize FB}}}^{l}$ versus $A^{\ell \ell}$ asymmetries compared to various predictions. The ellipses represent contours of total uncertainty at one, two, and three s.d. on the measured result by \dzero.}
 \end{SCfigure}

\begin{figure}[ht]
     \centerline{\includegraphics[width=0.975\columnwidth]{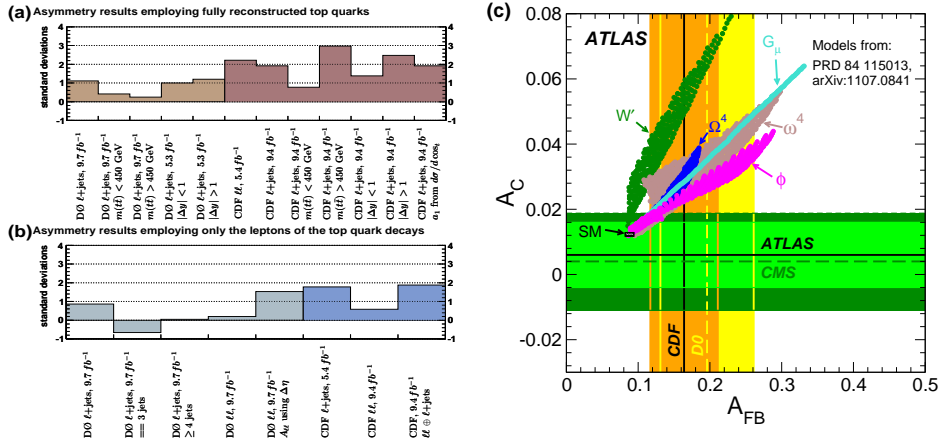}}
  \protect\caption{\label{fig:afb_ac_models} Summary (a) of the top quark and (b) lepton based asymmetries at the Tevatron, and (c) shows the inclusive top quark asymmetry results and their uncertainties from the Tevatron (vertical bands) and the LHC (horizontal bands) in the $A_{\mbox{{\footnotesize FB}}}$ versus $A_{\mbox{{\footnotesize C}}}$ plane \cite{atlas_afbSum}. Results are compared to predictions from the SM as well as predictions incorporating various potential new physics contributions.}
 \end{figure}

CDF uses data corresponding to $9.4~\mathrm{fb^{-1}}$ of integrated luminosity and employs a kinematic reconstruction to reconstruct the \ttbar final state in the \ljets decay channel. CDF measures an inclusive asymmetry of $0.164 \pm 0.045$ (stat. + syst.) at the parton level and also measures the kinematic dependency of $A_{\mbox{{\footnotesize FB}}}$, by measuring $\Delta y$ in bins of \mTT \cite{cdf_afb}. The dependency of $A_{\mbox{{\footnotesize FB}}}^{t\bar{t}}$ is shown in Fig.~\ref{fig:cdf_mttafb}(a).
The results show a dependency which is different from the SM expectation by 2.4 standard deviations.\\
\dzero uses the full Run II data corresponding to $9.7~\mm{fb^{-1}}$ of integrated luminosity and also fully reconstructs the \ttbar final state using a kinematic reconstruction. The measurement in the \ljets decay channel results in an inclusive asymmetry of $0.106 \pm 0.030$ (stat. + syst.) at the parton level \cite{d0_afb_update}. The result is compatible with the SM and results by CDF. It agrees within 0.3 standard deviations with the predictions ranging from 5\% (\mcatnlo) to 8.8\% (NLO QCD $\oplus$ EW corrections) \cite{bernSi}. \dzero does not see an indication for a strong \mTT dependency beyond the one expected by the SM as shown in Fig.~\ref{fig:cdf_mttafb}(b). It should be noted that predictions at NNLO pQCD are really needed to clarify the current picture in terms of deviations of existing $A_{\mbox{{\footnotesize FB}}}$ measurements versus the predictions. \\
CDF also measures the differential cross section for \ttbar production as a function of the top quark production angle $\cos \theta_t$ \cite{cdf_afb_costop}. Events are selected in the \ljets channel by requiring an isolated lepton with a $p_T$ of at least 20 GeV, \met$> 20$ GeV, at least three jets with 20 GeV and at least another loose jet with 12 GeV. Further quality cuts are applied. A complex characterization in terms of Legendre polynomials is carried out and compared to the SM prediction at NLO. The even moments have no contribution to the asymmetry, as the even Legendre polynomials are symmetric. The data agrees with the SM except for the first Legendre moment measured to $a_1= 0.40\pm0.09\,(\mm{stat.})\pm 0.08\,(\mm{syst})$ showing a two s.d. deviation to the SM expectation. This implies that the observed $A_{\mbox{{\footnotesize FB}}}$ is dominated by this first Legendre moment, which constrains possible contributions of new physics. \\
The \dzero result in terms of the lepton based asymmetries in the \ljets channel is $A_{\mbox{{\footnotesize FB}}}^{l} = 0.047 \pm 0.026$ (stat. + syst.) at the parton level \cite{d0_afb_lep} and in the dilepton channel the corresponding measurement is $A_{\mbox{{\footnotesize FB}}}^{l} = 0.044 \pm 0.039$ (stat. + syst.), whereas the dilepton asymmetry is measured to be $A^{\ell \ell} = 0.123 \pm 0.056$. It is interesting to note that the ratio of the two lepton based asymmetries in the dilepton channel shows a deviation from the SM prediction of about two standard deviations (see Fig.~\ref{fig:al_all_d0}).
Combinations within \dzero and across Tevatron are currently being investigated.\\
CDF employs data corresponding to up to $9.4~\mm{fb^{-1}}$ of integrated luminosity and measured $A_{\mbox{{\footnotesize FB}}}^{l} = 0.09 ^{+0.028}_{-0.026}$ after combining their results from \ljets and dilepton channels \cite{CDF-CONF-2013-11035}. The results discussed above are summarized in Fig.~\ref{fig:afb_ac_models}(a) employing the theoretical predictions for the specific quantity of $A_{\mbox{{\footnotesize FB}}} = 6.6 \pm 2 \%$ \cite{afbtheory} and $A_{\mbox{{\footnotesize FB}}}^{l} = 3.6 \pm 1 \%$ \cite{afbltheory}. Overall the experimental measurements are in excess of the SM prediction for $A_{\mbox{{\footnotesize FB}}}$ with CDF results showing larger deviations from the SM in various channels. For $A_{\mbox{{\footnotesize FB}}}^{l}$ experimental results deviate less from the predicted SM value. It should be noted that individual results on $A_{\mbox{{\footnotesize FB}}}^{l}$ employ the full data recorded by CDF and \dzero and a combination of $A_{\mbox{{\footnotesize FB}}}^{l}$ is currently ongoing.
A summary of the experimental results from the Tevatron and the LHC, which also shows the complementarity of the two machines together with the constraints on possible new physics contributions is provided in Fig.~\ref{fig:afb_ac_models}(c). Currently the experimental results provide a puzzling situation, as the Tevatron results show indications for deviations from the SM predictions by about one to two s.d., whereas the results from the LHC are not yet significant enough. Both communities are working hard to improve on the measurement techniques and the related uncertainties. A study performed in the context of the CSS 2013 \cite{snowmass} discusses the prospects of measurements of $A_{\mbox{{\footnotesize C}}}$ and their significances at the upcoming run of the LHC.

\section{Top Quark Mass}
\label{toc:mass}

\begin{figure}[ht]
    \centering
     \includegraphics[width=0.975\columnwidth]{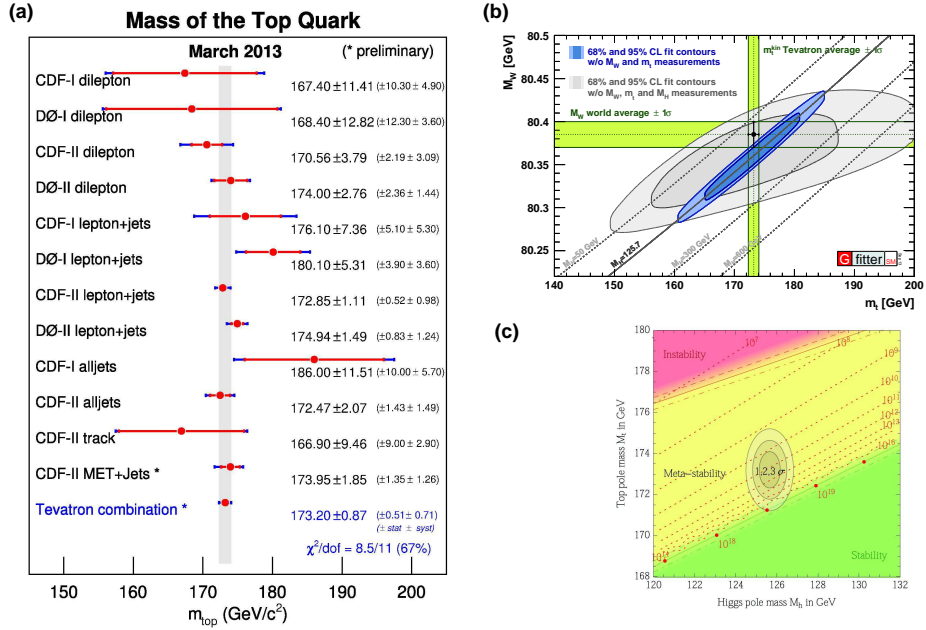}
  \protect\caption{\label{fig:mass_vsW} Overview (a) of all the measurements of the top quark mass by CDF and \dzero including the latest Tevatron combination \cite{tev_massCombi}. The (b) $m_t$ versus $m_W$ plane, note the scale in the case of $m_W$. The best measurements of $m_t$ (Tevatron) and $m_W$ (world average) and their one s.d.~uncertainties are indicated by the vertical and horizontal bands, respectively. The narrower blue and larger gray allowed regions indicate the results of the SM fit including and excluding the measurements of $m_H$, respectively \cite{gfitter}. The (c) regions of stability, meta-stability and instability of the SM vacuum in the $m_t$ -- $m_H$ plane \cite{vacuum}.}
 \end{figure}

A large number of measurements of the top quark mass have been carried out at the Tevatron and it is not possible to summarize all the details adequately in this short review. Instead the most precise measurements per decay channel performed by either CDF or \dzero are briefly summarized here, as well as the latest Tevatron combination. An overview of all the various measurements is given in Fig.~\ref{fig:mass_vsW}(a) including the result of the latest Tevatron combination with a precision of less than 0.5\% \cite{tev_massCombi}.\\
In case of \dzero the most precise measurement in the \ljets channel \cite{d0_mass} employs the so-called matrix element method (ME) which calculates an event probability density from differential cross sections and detector resolutions. The transfer-function relates the probability density of measured quantities to the partonic quantities. As one of the $W$ bosons decays hadronically a constraint on the $W$ mass can be used to fit the jet energy scale in-situ. The measurement uses $3.6~\mm{fb^{-1}}$ and is currently the most precise \dzero mass measurement across all decay channels. It yields a mass of $m_{t} = 174.9 \pm 0.8\, (\mm{stat.}) \pm 1.2\,(\mm{sys.\,+\,JES})~\mm{GeV}$.\\
The most precise measurement in the dilepton decay channel employs the neutrino weighting technique in  $5.4~\mm{fb^{-1}}$ of \dzero data, where the measurement integrates over the unknown neutrino momenta. This sample is limited in statistics due to the small $\cal{B}$ but has very low backgrounds. The in-situ JES correction is transferred from the \ljets measurement and reduces JES uncertainties. The measurement yields $m_{t} = 174.0 \pm 2.4 (\mm{stat.}) \pm 1.4 (\mm{sys.})~\mm{GeV}$ \cite{d0_mass_dilepton}. Measurements employing the full Run II data set are currently carried out by \dzero and results are expected very soon.\\
CDF has measured the top quark mass in the dilepton channel using the neutrino weighting and an alternate-variable template technique using 9.1 fb$^{-1}$ corresponding to the full Run II data set \cite{cdf_mass_dilepton}. The \ttbar dilepton final state is reconstructed by minimizing a $\chi^2$ function to scan over the space of possibilities for the azimuthal angles of neutrinos. It yields a mass of $m_{t}=170.80 \pm 1.83\,(\mm{stat.}) \pm 2.69\,(\mm{sys.})~\mm{GeV}$.\\
The most precise CDF measurement employs 8.7 fb$^{-1}$ in the \ljets decay channel and uses a template method \cite{cdf_mass_ljets}. Similar to the \dzero measurement in the \ljets channel also here the reconstructed $W$ boson provides an in-situ calibration for the JES reducing significantly the uncertainty originating from JES corrections. The template fit is carried out in sub-samples characterized by the number of $b$-tagged jets. The measurement yields $m_{t}=172.85 \pm 0.71\,(\mm{stat.\,+\,JES}) \pm 0.84\,(\mm{sys.})~\mm{GeV}$. CDF also measured the top quark mass in the all-jets decay channel \cite{cdf_mass_alljets} similarly employing the in-situ JES calibration as in case of the measurement in the \ljets channel. The measurement yields a top quark mass of $m_{t}=172.5 \pm 1.4\,(\mm{stat.}) \pm 1.0\,(\mm{JES}) \pm 1.1\,(\mm{sys.})~\mm{GeV}$.\\

Figure \ref{fig:mass_vsW}(b) shows the $m_t$ versus $m_W$ plane with the best measurements of $m_t$ (Tevatron combination) and $m_W$ (world average). Together with the measurement of the mass of the recently discovered Higgs boson \cite{higgs1, higgs2} this is a strong self-consistency test of the SM. Figure \ref{fig:mass_vsW}(c) shows regions of absolute stability, meta-stability and instability of the SM vacuum in the $m_t$ -- $m_H$ plane in the region of the preferred experimental range of $m_H$ and $m_t$ where the gray areas denote the allowed region at significances of one, two, and three standard deviations \cite{vacuum}. The current measurements and the theoretical extrapolation seem to indicate that the vacuum is meta-stable or, maybe even more peculiar, sitting right at the boundary of stability and meta-stability.

\section{Conclusions}

Recent measurements in the top quark sector at the Tevatron were discussed showing the complementarity of those to measurements at the LHC. More information about other Tevatron top quark measurements not discussed here can be found in Refs.~\cite{Tevwebpages,Tevwebpages2}. Exciting new results were recently presented by the Tevatron experiments: the observation of $s$-channel single top quark production, which is very hard to measure in LHC conditions and a variety of measurements targeted at the still puzzling situation observed in measurements of asymmetries in the top quark sector. Additonal information on this situation is provided by precise measurements of differential cross section distributions. In case of \dzero measurements it should be noted that a variety of measurements are currently updated to the full Run II data set and results are expected very soon. This includes measurements of the top quark mass, where together with additional studies on systematic uncertainties a significant improvement of the experimental uncertainties is expected.\\
The Tevatron continues to provide exciting measurements, which are competitive and complementary to measurements at the LHC. The combination of CDF and \dzero measurements is part of the legacy of the Tevatron, which is being written right now and will stay for a long time in the text books.

\section*{Acknowledgments}
The author thanks B.~Hirosky, R.~v.~Kooten, S.~Shary and J.~Wilson for helpful comments when putting together this review article.



\end{document}